\documentclass{camera}
\usepackage{graphicx}  

\def\lesssim{\mathrel{\hbox{\rlap{\hbox{\lower4pt\hbox{$\sim$}}}\hbox{$<$}}}}
\def\gtrsim{\mathrel{\hbox{\rlap{\hbox{\lower4pt\hbox{$\sim$}}}\hbox{$>$}}}}

\begin{document}

%
\title{PUZZLING AFTERGLOW's OSCILLATIONS IN GRBs AND SGRs:
 TAILS OF PRECESSING  JETS}

%


\author{D.Fargion}
%
\organization{Physics department, Universita' degli studi "La Sapienza", \\
              Piazzale Aldo Moro 5, -  00185 Roma, Italy\\
              INFN Roma 1, Rome , Italy}

\maketitle

%

\begin{abstract}
Damped  oscillating afterglows in GRB $030329$ and in SGR $1900
+14$ find a natural explanation in precessing $\gamma$ Jet model
for both GRBs and SGRs. The very thin Jet cone
$\frac{\Delta\Omega}{\Omega}\leq 10^{-8}$ combines at once the
Supernova power and the apparent huge GRBs output:
$\dot{E}_{GRBs} \simeq$
$\dot{E}_{SN}$$\frac{\Omega}{\Delta\Omega}$ leading to a  better
understanding of their remarkable GRB-Super Nova connection shown
in early GRB$980425$/SN$1998$bw event and in most recent
GRB$030329$/SN$2003$dh  one.  The same  thin beaming offer a
understanding of the apparent SGR-Pulsar power connection:
$\dot{E}_{SGRs} \simeq$
$\dot{E}_{Xpuls}$$\frac{\Omega}{\Delta\Omega}$. The precessing
Jet model for both GRBs and SGRs, at their different luminosity,
explains the existence of a few identical energy spectra and time
evolution of these two sources leading to their unified
understanding. The spinning-precessing Jet explains the rare
mysterious X-Ray precursors in GRBs and SGRs. The
Multi-precessing Jet at peak activity in all band may explain the
puzzling X or optical re-brightening bumps found in last
GRB$030329$ and earliest SGR $1900+ 14$ on $27$ August $1998$ and
on $18$ April $2001$, as well as the multi-bump radio light-curve
observed in GRB$980425$ and  GRB$030329$. Rarest micro-quasars in
our galaxy as SS433, and Herbig Haro objects describe these thin
precessing Jet imprint in their $3$D relic nebulae shapes.
\end{abstract}
\section{INTRODUCTION}
The very clear imprint of gamma polarization in the $\gamma$
signals from GRB$021206$ \cite{Coburn} probes to the eyes of most
skeptical Fireball theorist the very presence of a thin
collimated jet (opening angle $\Delta\theta \leq 0.6^o$;
$\frac{\Delta\Omega}{\Omega}\leq 2.5 \cdot 10^{-5} $) in Gamma Ray
Burst, GRBs. The very last and well proved GRB$030329$/SN$2003$dh
time and space coincidence confirms definitively  the earliest
GRB$980425$/SN$1998$bw connection among GRB and Supernova.
Therefore the apparent  extreme GRBs  luminosity is just the
extreme beamed blazing gamma Jet observed in axis during a
Supernova event. Indeed (or moreover) the maximal isotropic SN
power, $\dot{E}_{SN}$$ \simeq 10^{45}$ erg $s^{-1}$, because of
very probable energy equi-partition, should be collimated even
into a more thinner jet $\frac{\Delta\Omega}{\Omega}\leq
10^{-8}$, in order to explain at once the apparent observed
maximal GRBs output, $\dot{E}_{GRB}$$ \simeq 10^{53}$ erg
$s^{-1}$. Consequently one-shoot thin Jet  GRBs needs many more
$\dot{N}_{GRBs} \simeq$ $\frac{\Omega}{\Delta\Omega}\geq 10^{8}$
events than any spread isotropic Fireballs. Such a rate exceed
 even the known Supernova one. To overcome the puzzle a persistent precessing
decaying  jet (life-time $\tau_{Jet}$ $ \geq 10^3$ $\tau_{GRB}$)
is compelling. Relic GRBs  sources may be found in compact SNRs
core, as  NS or BH jets;  at later epoch, their lower power
$\gamma$ jets may be within detectability only from nearby
galactic distances, as Soft Gamma Repeaters (SGRs) or anomalous
X-ray Pulsars, AXPs. This common Jet nature explain some
connection between GRBs and SGRs.  Indeed rare spectra of SGRs
behaved as GRBs. Also X-Ray precursors in GRBs as well as in SGRs
suggest the needed  for a precessing Jet model. A surprising multi
re-brightening afterglows observed in early and late GRB $030329$
optical transient, like in the $27$ August $1998$ and $18$ April
$2001$ SGR $1900+14$ events, might be  the damped oscillatory
imprint left by such a multi-precessing $\gamma$-X-Optical and
Radio  Jet.
\subsection{The different GRBs puzzles} Gamma Ray Burst mystery
lays  in its huge energy fluence, sharp variability, extreme
cosmic distances and very different morphology.  A huge isotropic
explosion (the so called Fireball) was the ruling  wisdom all
along last decade.  However shortest millisecond time
 scales called for small compact objects, so contained and confined
 to became opaque, because abundant pair production, to their own explosive luminosity  (over Eddington luminosity) and
 so small in size and in masses (few solar masses) to be  unable to supply themselves the needed
 larger and larger isotropic energies. The spectra , in a Fireball, had to be nearly thermal, contrary to data.
 Fireball became  shell by shell  an hybrid complex
 model, where  power law after power law,
it tried to fit each complex GRBs spectra and time evolution.
 The huge GRBs powers as GRB$990123$ made the final
  collapse of the Fireball  model. New  families of Fire-Ball Jet
  (and their label names like Hyper-Nova, Supra-Nova, Collapsar )
  models  alleviated, by a Jet beaming, the explosive energy budget request.
   However in this compromise attitude the puzzle of the
   GRB$980425$/SN$1998$bw  (which require a very thin Jet observed off-axis \cite{bib7},\cite{bib11quater}) has been cured by most  skeptical authors
   by a tenacious cover-up (neglecting or refuting The GRB-SN  existence)
   or claiming the co-existence of an new  zoo of GRBs \cite{Bloom2000},\cite{kulkarni}.
   These new compromised fountain-like  Fireball model has been collimated in a Jet
 within a $10^o$ angle beam, as a soft link  between past Fireball and emerging
 Jet. However the apparent required GRB power output is still huge ($10^{50}$ erg $s^{-1}$),
  nearly $10^5$ more intense
 than other known maximal explosion power (the Super-Nova one).
  More and more evidences in last years and more recently have shown that
  Super-Nova might  harbor a collimated Jet Gamma Ray Burst
   (GRB$980425$/SN$1998$bw ,GRB$030329$/SN$2003$dh).
   To combine the Super-Nova Luminosity and the apparent huge GRBs
   power one need a very much thinner beam jet, as small as a solid angle $\Delta\Omega/\Omega \simeq$ $10^{-7}$
   or   $10^{-8}$ respect to  $\Omega \simeq 4 \pi$,(corresponding to a Jet  angle  $0.065^o-0.02^o$ ).
    There is a statistical need \cite{bib11quater}
   to increase the GRB rate inversely to the beam Jet solid angle.
   The needed SN rate (to explain GRBs) may even exceed the observed one
   (at least SN type Ib andIc) event in our
    observable Universe ($\dot{N_{NS}}$ $\leq 30
   s^{-1}$). Indeed assuming that only a  fraction of the SN
   (with optimistic attitude $0.1$ of all known SN) experience an asymmetric Jet-SN
   explosion, than the  corresponding observed rate  $\dot{N_{GRBs}}$ $\simeq  10^{-5} s^{-1}$ and
   $\dot{N_{SN}}$ $\simeq  3 s^{-1}$  imply
   $\frac{\dot{N_{GRBs}}}{\frac{\Delta\Omega}{\Omega}} \simeq 10^{2} s^{-1} \longleftrightarrow 10^{3} s^{-1}$
   a result  nearly $2-3$  order of magnitude larger  than the observed SN rate.
   In this frame one  must assume a GRB  Jet with a continuous active, decaying  life-time
  much larger than GRBs duration itself at least by a corresponding scale $\tau_{Jet} \simeq 10^{3}\tau_{GRBs}$.
   Indeed we  considered GRBs  (as well as Soft Gamma Repeaters SGR) as
  very thin blazing ($\leq 0.1^o$) precessing Gamma Jets
  spinning and precessing \cite{bib8},\cite{bib9},\cite{bib12},\cite{bib16bis},\cite{bib11quater},\cite{Fargion2001a};
  in this scenario  GRBs are born within a Super-Nova
  power collimated inside a very thin beam  able to blaze  us by an apparent GRB intensities.
  The inner angle geometrical dynamics  while spinning and precessing induce
  the wide $\gamma$ burst  variability able to fit  the very different observed GRBs ones.

\begin{figure}\centering\includegraphics[width=8cm]{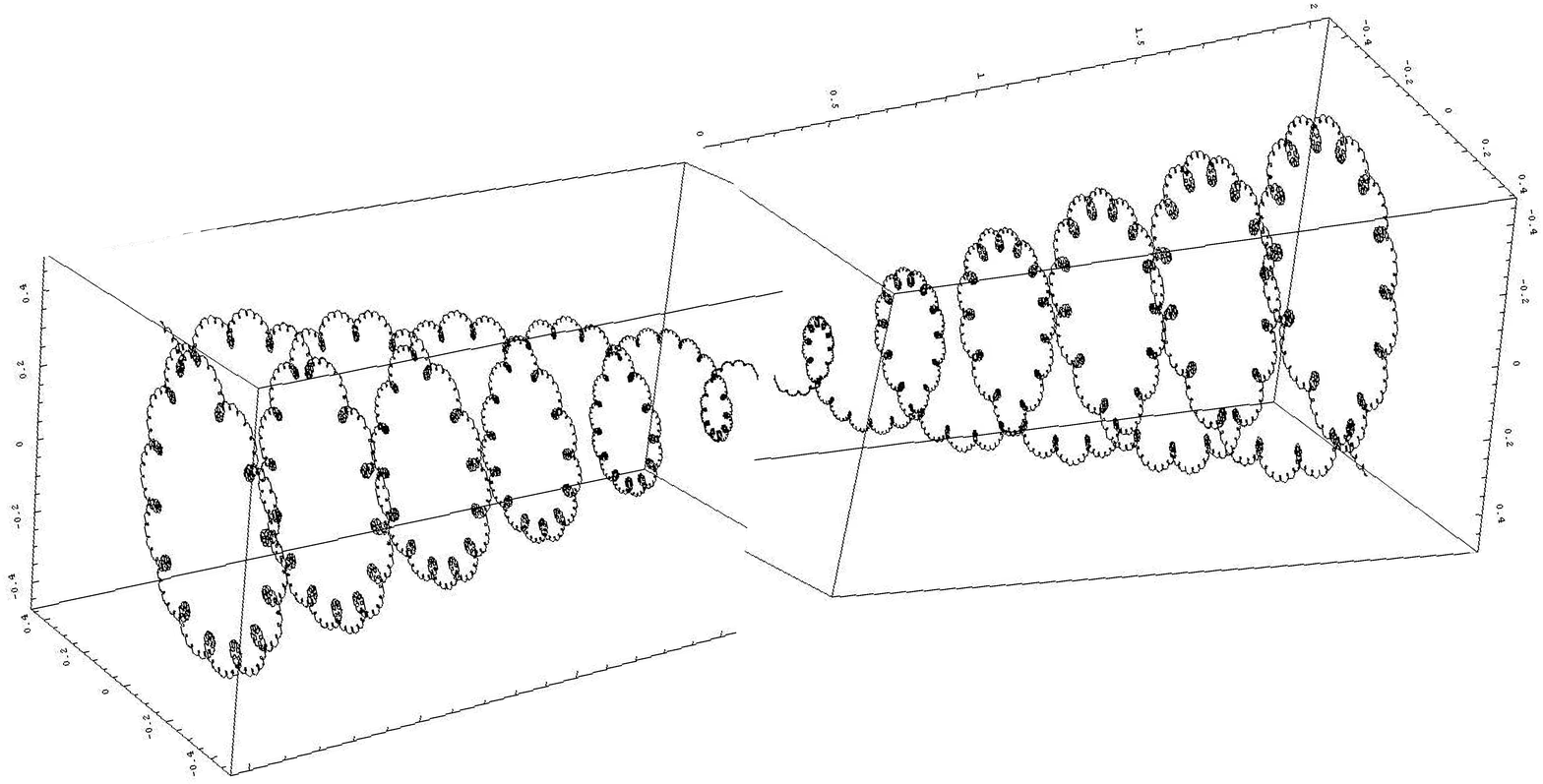}
\caption {A possible inner structure 3D of a multi precessing Jet;
its cone structures and its  stability at late stage it maybe
reflected in the quasi-periodic repetitions of the  Soft Gamma
Repeaters while beaming to us along the cone edges  toward the
observer. Its early blast  at maximal SN out-put may simulate a
brief blazing GRBs event, while a fast decay (hours scale) may
hide its detectability below the threshold, avoiding in general
any common  GRB  repeater.} \label{fig:fig1}
\end{figure}
   The averaged $\gamma$ jet deflection from the   axis of sight defines a main early power law
decay; an inner damped oscillatory substructure
   may be observed, as the amazing damping oscillatory afterglows in GRB$030329$.  The very thin and
collimated and  long life decaying jet (opening angle $\theta
\leq 0.05^o$, whose decay power law life-time of a few hours
occurs with an exponent $\alpha \simeq -1$), while spinning and
precessing at different scale times, it may trace and may better
explain the wobbling of the $\gamma$ GRBs  and the long train of
damped oscillations of the X tail afterglows   within hours, the
optical transient during days and weeks later. The GRBs
re-brightening are no longer a mystery as in a one-shoot model.
These wobbling signatures may be also be found in rarest and most
powerful and studied SGRs events. The spread and wide conical
shape of these precessing twin jets may be recognized in a few
relic SNRs as in  the twin SN 1987A wide external  rings, the
Vela arcs and the spectacular Egg Nebula dynamical shape.
\begin{figure}\centering\includegraphics[width=8cm]{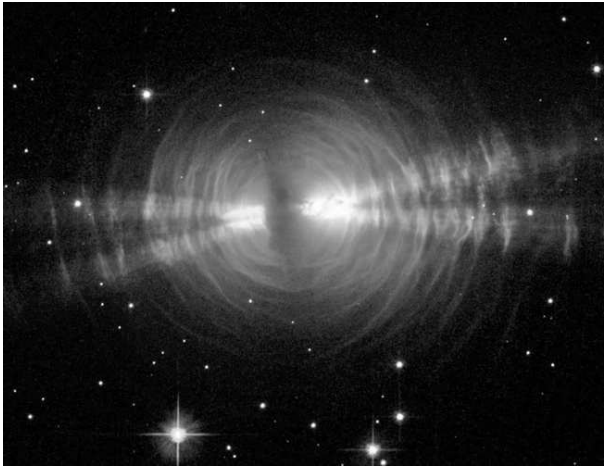}
\caption {Egg Nebula Twin Conical Shape. Such a mysterious twin
cone picture maybe easily understood within a precessing Jet
source well hidden in the center of the nebula whose presence is
made evident by the help of a diffused  nearby  cloud target}
\end{figure}

\subsection{The  geometrical multi-precessing Gamma Jet in GRB }
 We imagine the GRB and SGR
nature as the early and the late stages of jets fueled first by
SN event and later by an asymmetric accretion disk or a companion
(WD, NS) star.


Their binary angular velocity $\omega_b$ reflects the beam
evolution $\theta_1(t) = \sqrt{\theta_{1 m}^2 + (\omega_b t)^2}$
or more generally a multi-precessing angle $\theta_1(t)$
\cite{bib9},\cite{bib10}:
$\theta_1(t)=\sqrt{\theta_{x}^2+\theta_{y}^2 } $
$$
  \theta_{x}(t) =                               
  \theta_{b} sin(\omega_{b} t+ \varphi_{b} )+
  \theta_{psr}sin(\omega_{psr} t +  \varphi_{psr})+
  \theta_{N}sin(\omega_{N} t  + \varphi_{N})
$$

\begin{equation}
  \theta_{y}(t) = \theta_{1 m}+
  \theta_{b} cos(\omega_{b} t + \varphi_{b})+
  \theta_{psr} cos(\omega_{psr} t +  \varphi_{psr})+
  \theta_{N} cos(\omega_{N} t  + \varphi_{N})
\end{equation}
where $\theta_{1 m}$ is the minimal impact angle parameter of the
jet toward the observer, $\theta_{b}$, $\theta_{psr}$,
$\theta_{N}$ are, in the order, the maximal opening precessing
angles due to the binary, spinning pulsar, nutation mode of the
multi-precessing jet axis. The arbitrary phase $ \varphi_{b}$, $
\varphi_{psr}$, $\varphi_{N}$, for the binary, spinning pulsar
and nutation,  are able to fit the complicated GRBs flux
evolution in most GRB  event scenario. Naturally it is very
possible to enlarge the parameter to a fourth precession angular
component whose presence may better fit the wide spread of scale
variability; here we shall constrains to a three parameter
precession  beam.

\begin{figure}\centering\includegraphics[width=8cm]{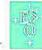}
\caption {A whip  behaviour of HH34 micro-quasar narrow Jet. The
long tail of the jet in both near by and far side describes a thin
moving jet; an internal  spinning sub-structure may be hidden
inside the strip jet width.}
\end{figure}
\begin{figure}\centering\includegraphics[width=8cm]{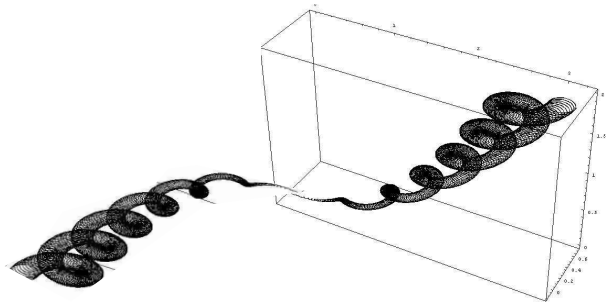}
\caption {A twin spinning and precessing jet configuration whose
 appearance on line of sight, when at maximal supernova
output, may blaze as the sudden GRBs; at much later and less
powerful stages, while behaving as an X-Ray pulsar, it may rarely
blaze as a SGRs from nearby, galactic like distances.}\end{figure}

\begin{figure}\centering\includegraphics[width=8cm]{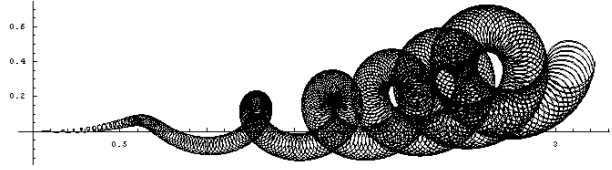}
\caption {The consequent twin spinning and precessing jet
configuration projected onto a $2$dimensional screen whose blazing
appearance on line of sight, at SN may appear as the sudden GRBs
or later at less powerful stages  as a SGRs from nearby
distances.}
\end{figure}
\begin{figure}\centering\includegraphics[width=8cm]{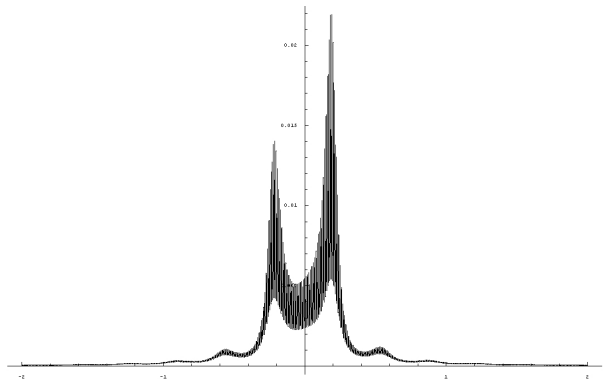} 
\caption{ Multi bump afterglow behaviour of the intense
precessing Jet above whose blazing shows the characteristic
oscillatory damped decay as the recent GRB $030329$ and the
intense SGR on $27$ August 1998. The luminosity starting time is
assumed near zero ( at SN birth time). In present simulation the
assumed Lorents factor is $\gamma_e$$= 2 \cdot 10^3$ }
\end{figure}
\begin{figure}\centering\includegraphics[width=8cm]{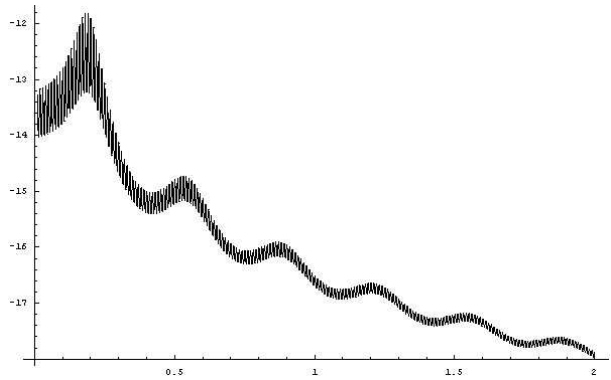} 
\caption{ Multi bump Flux Intensity behaviour in linear scales,
normalized to visual magnitude for a previous precessing Jet
simulating the characteristic oscillatory damped decay as the
recent GRB $030329$ and the intense SGR on $27$ August 1998; time
scale are arbitrary; in the GRB $030329$ the unity corresponds to
nearly day scale while in SGR event the unity in much smaller
minute scale}
\end{figure}
\begin{figure}\centering\includegraphics[width=8cm]{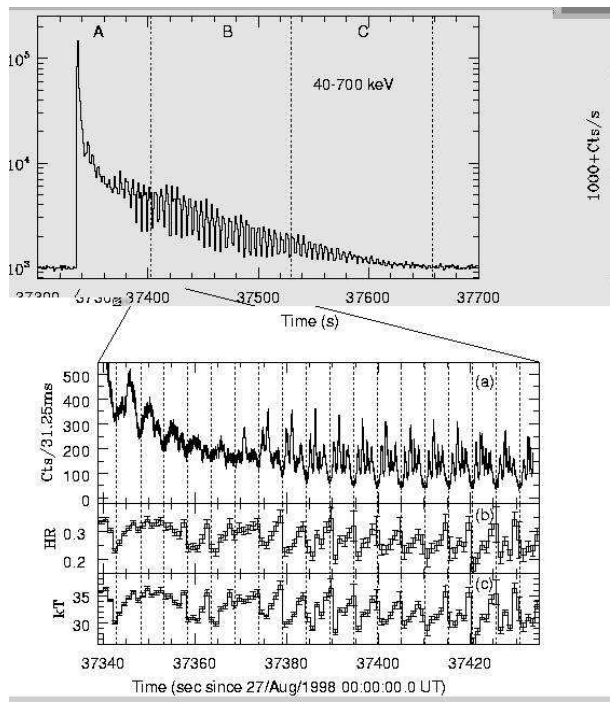} 
\caption[h]{ Multi bump behaviour of the intense SGR on $27$
August 1998, showing the characteristic oscillatory damped decay
as GRB $030329$ described above}
\end{figure}
\begin{figure}\centering\includegraphics[width=8cm]{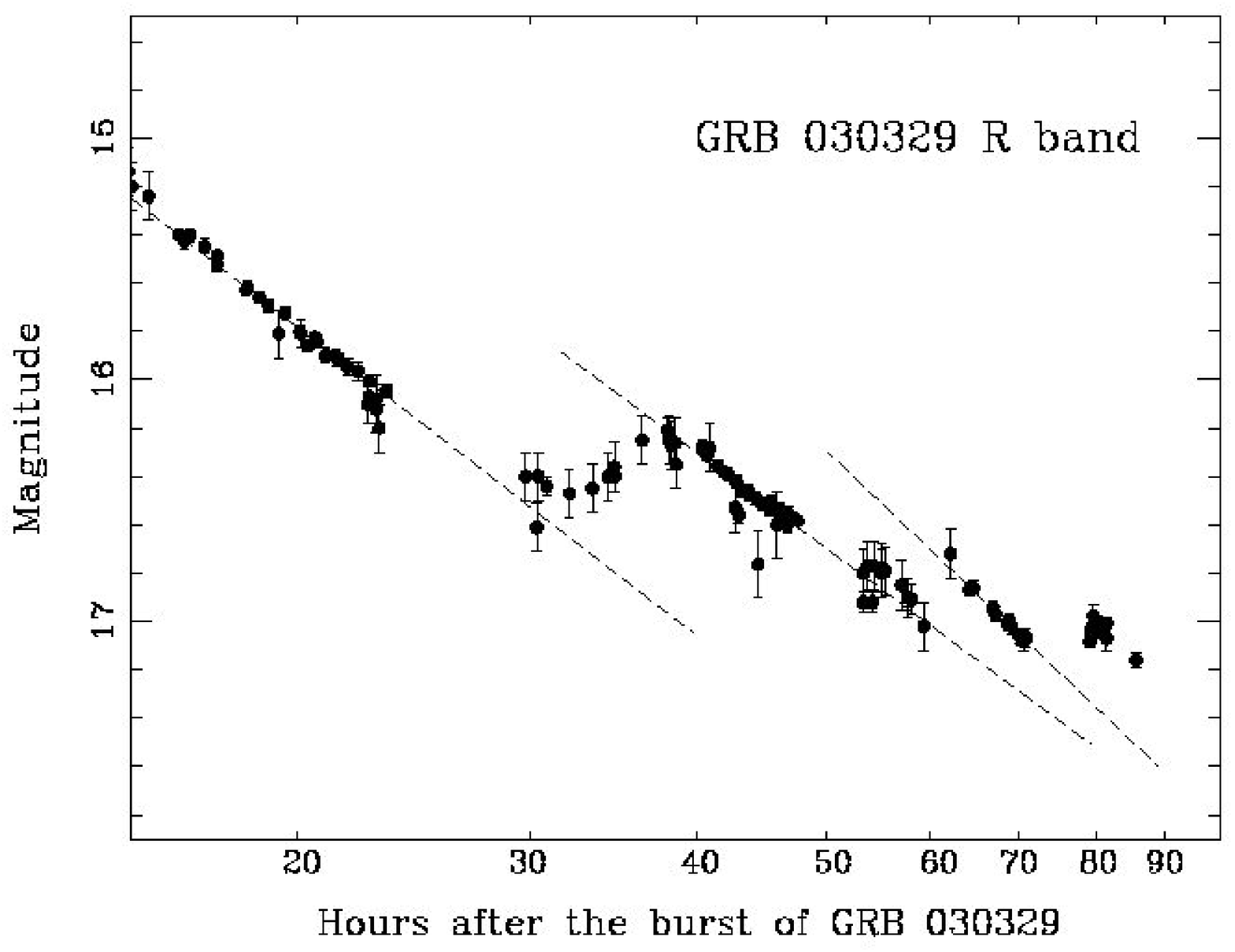} 
\caption {The multi bump behaviour or re-brightening of the
oscillatory damped decay observed in GRB $030329$; its puzzling
imprint maybe described by a precessing $\gamma$, $X$ and optical
jet }
\end{figure}

    For a 3D pattern and its projection along the vertical
axis in an orthogonal 2D plane see following descriptive pictures.
The different  angular velocities are combined in the
multi-precession wobbling. Each bending component  is keeping
memory either of the pulsar jet spin angular velocity
($\omega_{psr}$) and its opening angle $\theta_{psr}$, its
nutation speed ($\omega_N$) and nutation angle $\theta_{N}$ (due
to possible inertial momentum anisotropies or beam-accretion disk
torques); a slower  precession  by the binary $\omega_b$
companion (and its corresponding open angle $\theta_{b}$) will
modulate the overall jet precession. On average, from eq.(3) the
$\gamma$ flux and the $X$, optical afterglow are decaying on time
as $t^{-\alpha}$ , where $\alpha \simeq 1-2$; however the more
complicated spinning and precessing jet blazing is responsible
for inner small scales wide morphology of GRBs and SGRs as well
as their partial internal periodicity. The consequent $\gamma$
time evolution and spectra derived in this ideal models may be
compared successfully with observed GRB data evolution.
\section{Hard $\gamma$-X  Jet by Inverse Compton Scattering }
The $\gamma$ Jet is born mainly by Inverse Compton Scattering by
GeVs electron pairs onto thermal photons \cite{bib8}
\cite{bib12}, \cite{bib16},\cite{bib17}, \cite{bib11quater} in
nearly vacuum space. Therefore these   electron pairs are boosted
in the Jet at Lorentz factor $\gamma_e \geq 2 \cdot 10^3$.
   Their consequent Inverse Compton Scattering will induce a parallel $\gamma$ Jet
   whose beam angle is $\Delta \theta \leq \frac{1}{\gamma} \simeq 5\cdot10^{-4} rad \simeq 0.0285^o $
   and a wider, less collimated X, Optical cone. These beaming
   angles agree with the one assumed to explain the required beamed GRBs-SN
   powers.   Indeed  the electron pair Jet may generate  a secondary beamed synchrotron
   radiation component at radio energies, in analogy to the behaviour
    of  BL Lac blazars whose hardest TeV $\gamma$ component is made by Inverse Compton Scattering
    while its  correlated  X band emission is due to the synchrotron component. Anyway the inner jet
   is dominated by harder photons while the external cone contains softer $X$,
   optical and radio waves. A jet angle related by a relativistic kinematics would imply $\theta \sim
\frac{1}{\gamma_e}$, where $\gamma_e$ is found to reach $\gamma_e
\simeq 10^3 \div 10^4$ \cite{bib11quater},\cite{bib10}. At first
approximation the gamma constrains is given by Inverse Compton
relation: $< \epsilon_\gamma > \simeq \gamma_e^2 \, k T$ for $kT
\simeq 10^{-3}-10^{-1}\, eV$ and $E_e \sim GeVs$ leading to
characteristic X-$\gamma$ GRB spectra.  The origin of $GeVs$
electron pairs are  very probably decayed secondary related to
primary inner muon pairs jets, able to cross dense stellar target
\cite{bib17} . The consequent adimensional photon number rate
 as a function of the observational angle
$\theta_1$ responsible for peak luminosity  becomes
\cite{bib11quater}
\begin{equation}
\frac{\left( \frac{dN_{1}}{dt_{1}\, d\theta _{1}}\right) _{\theta
_{1}(t)}}{ \left( \frac{dN_{1}}{dt_{1}\, d\theta _{1}}\right)
_{\theta _{1}=0}}\simeq \frac{1+\gamma ^{4}\, \theta
_{1}^{4}(t)}{[1+\gamma ^{2}\, \theta _{1}^{2}(t)]^{4}}\, \theta
_{1}\approx \frac{1}{(\theta _{1})^{3}} \;\;.\label{eq4}
\end{equation}
The total fluence at minimal impact angle $\theta_{1 m}$
responsible for the average luminosity  is
$$
\frac{dN_{1}}{dt_{1}}(\theta _{1m})\simeq \int_{\theta
_{1m}}^{\infty }\frac{ 1+\gamma ^{4}\, \theta _{1}^{4}}{[1+\gamma
^{2}\, \theta _{1}^{2}]^{4}} \, \theta _{1}\, d\theta _{1}\simeq
\frac{1}{(\, \theta _{1m})^{2}}\;\;\;.
$$
These spectra fit GRBs observed ones
\cite{bib10},\cite{bib12},\cite{bib16}, \cite{bib11quater}.
Assuming a beam jet intensity $I_1$ comparable with maximal SN
luminosity, $I_1 \simeq 10^{45}\;erg\, s^{-1}$, and replacing
this value in the above a-dimensional equation  we find a maximal
apparent GRB power for beaming angles $10^{-3} \div 3\times
10^{-5}$, $P \simeq 4 \pi I_1 \theta^{-2} \simeq 10^{52} \div
10^{55} erg \, s^{-1}$, just within observed ones. We also
assumed a power law jet time decay as follows
$$
  I_{jet} = I_1 \left(\frac{t}{t_0} \right)^{-\alpha} \simeq
  10^{45} \left(\frac{t}{3 \cdot 10^4 s} \right)^{-1} \; erg \,
  s^{-1}
$$
where ($\alpha \simeq 1$) is a value able to reach, at 1000 years
time scales, the present known galactic micro-jet (as SS433)
intensities powers: $I_{jet} \simeq 10^{39}\;erg\, s^{-1}$. This
offer a natural link between the GRB and the SGR out-put powers.
We used the model to evaluate if April precessing jet might hit us
once again. It should be noted that a steady angular velocity
would imply an intensity variability ($I \sim \theta^{-2} \sim
t^{-2}$) corresponding to some of the earliest afterglow decay
law. These predictions have been proposed since a long time,
\cite{bib11quater}. Similar descriptions with more parameters and
within a sharp time evolution of the jet has been  also proposed
by other authors \cite{bib14},\cite{b35}.
\subsection{Precessing Radio Jet by Synchrotron Radiation}
The same GeVs electron pair Jet may generate  a secondary beamed
synchrotron   radiation component at radio energies, in analogy
to the behaviour    of  BL Lac blazars whose hardest TeV $\gamma$
component is made by Inverse Compton Scattering
    while its  correlated  X band emission is due to the synchrotron component. Anyway the inner jet
   is dominated by harder photons while the external cone contains softer $X$,
   optical and radio waves. Their precessing in wide angle is the
   source of the radio bumps at days scale times
   clearly observed recently in GRB$980425$,
   GRB$030329$ light curves.  The very peculiar oscillating GRB$970508$
    optical variability did show a re-brightening nearly  two months later, did also
    show a remarkable multi-bump variability in radio wave-lenght.  For this reason and
    we are more inclined to believe that this fluctuations were indeed to be related to the Jet
    precession and not to any interstellar scintillation. There is not any direct
   correlation between the $\gamma$ Jet made up by I.C. scattering and the Radio Jet because
   the latter is dominated by the external magnetic field energy density: there
   maybe a different beaming opening and a consequent different time modulation
   respect to the inner $\gamma$ Jet. However the present wide
   energy power emission between SN2002ap and GRB$030329$ radio light curves
   makes probable a beaming angle comparable: $\leq 10^{-3}- 10^{-4}$ radiant.


\section{ X Ray precursor by Precessing Jet }
\begin{figure}
\centering\includegraphics[width=10cm]{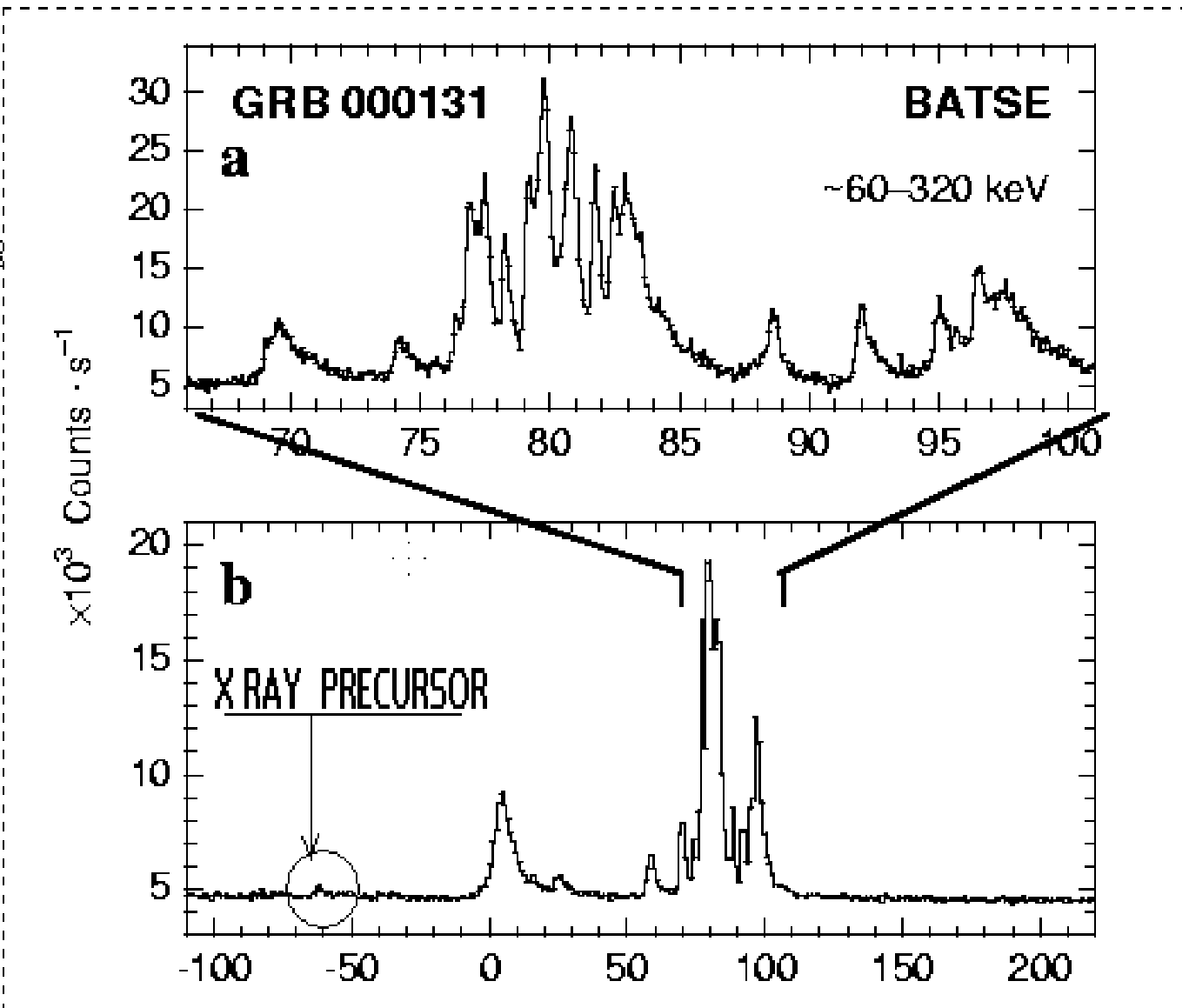}{X
precursors in  most distant known GRB, red-shifted $z= 4.5$:
GRB$000131$.}
\end{figure}

     The thinner the jet, the wider the sample and the source volume
    (largest redshift) and the harder is the $\gamma$ Jet
    observed. This explain why, in opposite behaviour respect
    to the Hubble cosmic expansion and to the time dilution, the
    observed    hardest GRBs spectra and the most variable and the most
    power-full ones are not the nearest but the most distant ones.
    Isotropic explosions are out of this frame. Indeed
   the extreme $\gamma$ energy budget, calling for a
   comparable $\nu$ one, exceeds  few or many solar masses in
   isotropic models  emission even for ideal full mass-energy conversion.
   To judge the fitting of the  Jet  model  let us  consider
   the  most distant $z= 4.5$ known GRB event:
   GRB$000131$ and its $X$ ray precursor.
   This event while being red-shifted and slowed down by a
   factor $5.5$ exhibited on the contrary shows a short scale time and a very fine structure
   in disagreement with any fireball model, but  well compatible  with
    a thin, fast spinning precessing $\gamma$ jet.
    Indeed let us notice the presence of a weak $X$-ray precursor pulse
   lasting 7 sec, 62 sec before the huge main structured $\gamma$ burst
   trigger  GRB$000131$ \cite{Fargion2001a}. Its arrival direction (within 12 degree error) with main GRB
   is consistent only with the main pulse (a probability to occur by chance below
   $3.6\cdot 10^{-3}$).   The time clustering proximity (one minute over a day GRB rate average) has the probability
   to occur by chance below once over a thousand.  The over all probability to observe this precursor by change is below 3.4 over a million
   making inseparable its association with the main  GRB000131 event.
    This weak burst signal correspond to a power above a million
   Supernova and have left no trace or Optical/X transient just a minute
   before the real (peak power $> billion $ Supernova) energetic event.
   No isotropic GRB explosive progenitor could survive such a
   disruptive   isotropic (million supernova output)  precursor
   trigger nor any multi-exploding jet.
   Only a persistent, pre-existing  precessing Gamma Jet
    crossing twice nearby the observer direction  could naturally explain this luminosity evolution.
   These X-ray precursor are not  unique but are found in $3-6\%$
   of all GRBs. Similar X precursors occurred in SGRs event
   as the $1900 + 14$ on 29 August 1998.






\section{Conclusions: Neutrino-Muon Jets Progenitors }
 We  believe that GRBs and SGRs  are persistent blazing flashes from light-house
thin $\gamma$ Jet spinning  in multi-precessing (binary,
precession, nutation)  mode. These GRBs Jets are originated by NSs
or BH in binary system or disk powered by infall matter; their
relics (or they progenitors) are nearly steady X-ray Pulsars
whose fast blazing is source of SGRs. The Jet is not a single
explosive event even in GRB, but they are powered at maximal
output during a long period of SN event. The beamed Jet power is
comparable to SN ones at its peak luminosity; this external
$\gamma$ Jet has a chain of progenitor identities: it is born in
most SN and or BH birth and it is very probably originated by a
very collimated inner primary muon Jet pairs at TeVs-Pevs
energies. These muons could cross with negligible absorption the
dense target lights along the SN explosions, nearly transparent to
photon-photon opacities. We speculate that these muon pair
progenitors might be themselves secondary relics beamed by
ultra-high energy neutrino Jet originated  in the very interior
of the new born  NS or BH, neutrino able to escape quite dense
matter envelopes obscuring the Super-Nova volume
\cite{Gupta},\cite{Fargion2002}. The high energy relativistic
muons (tens TeVs-PeVs energies) decay in flight in electron pairs
where the baryon density is still negligible; these muons are
source, by decay in flight to Tevs-GeV electron pair showering
whose final Inverse Compton Scattering with nearby thermal photon
is the final primary of the observed hard $X$ - $\gamma$ Jet. The
cost of this long chain of reactions  is a  poor energy
conversion, but the benefit is the possibility to explain the
$\gamma$ escape from a very dense explosive and polluted (by
matter and radiation) narrow  volume. The relativistic morphology
of the Jet and its multi-precession geometry is the source of the
complex $X$-$\gamma$ spectra signature of GRBs and SGRs. Its
inner Jet ruled by relativistic Inverse Compton Scattering, has
the  hardest and rarest beamed GeVs-MeVs photons (as the rare and
long $5000$ s  life  EGRET GRB$940217$ one) but its external Jet
cones are dressed by softer and softer photons. This   onion like
multi Jets is not totally axis symmetric: it may show itself
 while spinning and turning and spraying around in a
deformed (often and in particular in fast GRBs) elliptical onion
off-axis  rings, out-of-center; therefore ones their internal
harder  core at the extreme edges are leading to a most common
hard to soft GRBs $\gamma-X$ train signal. Nevertheless this  is
not the rule. In our present model and simulation this internal
effect has been  neglected without any major consequence. The
complex variability of GRBs and SGRs are simulated successfully
by the equations above
\cite{bib12},\cite{bib16bis},\cite{bib11quater}; the consequent
geometrical beamed Jet blazing may lead also to the observed
widest morphology $X-\gamma$ signatures.The mystery therefore is
   not longer in an apparent huge GRB luminosity, but in an extreme beam jet
   collimation and precession.




%
\end{document}